\documentclass[12pt]{article}
\usepackage{graphicx}
\usepackage{amsmath}
\usepackage{amssymb}
\usepackage{units}
  
\begin{document}

\title{Bayesian Optimisation of Non-linear Breit-Wheeler Pair Production in Simulated Laser Experiments}

\author{Christopher Arran\\
\small{Physics Department, Lancaster University, Bailrigg, Lancaster, UK, LA1 4YW}\\
\small{Cockcroft Institute, Keckwick Ln, Daresbury, Warrington, UK, WA4 4AD}\\
\small{c.arran@lancaster.ac.uk}\\
Stuart Morris\\
\small{Department of Physics, University of Warwick, Coventry, UK, CV4 7AL}\\
Christopher P. Ridgers\\
\small{York Plasma Institute, University of York, Heslington, York, UK, YO10 5DD}}
\vspace{10pt}

\maketitle

\begin{abstract}
High laser intensities enable the production of electron-positron pairs from bright gamma rays passing through strong fields. Potentially the most promising approach for all-optical experiments in the near term uses dense but higher divergence electron beams from laser wakefield acceleration to produce gamma rays through inverse Compton scattering. Achieving many-photon collisions between these gamma rays and the high intensity laser pulse in practice is extremely difficult, however, due to significant shot-to-shot jitter in laser pointing and timing. In practice, this jitter reduces the yield of electron-positron pairs by orders of magnitude.

We model these practical difficulties using simulated Monte-Carlo experiments. By using a more efficient algorithm for sampling infrequent pair production with particle splitting, we enable the exploration of a multi-dimensional parameter space. Using Gaussian Process Regression we then efficiently find optimal conditions for maximising pair production by changing the laser spot size, the energy in the colliding beam, and the stand-off distance between the laser wakefield accelerator and the focus of the colliding laser pulse. We find that the optimal stand-off distance increases with the degree of laser jitter and that the best conditions for producing electron-positron pairs are not the same as the best conditions for maximising the energy in the gamma rays. With \unit[100]{J} of laser energy, we estimate realistic rates of pair production of around 1 pair per 100 electrons are achievable even with jitter of 10s of microns and 10s of femtoseconds.
\end{abstract}

%
%
%
%
%

\section{Introduction}

Laser facilities continue to push the intensity frontier to new records, in recent years achieving peak powers of $\unit[10]{PW}$\cite{radier2022} and peak intensities of $\unit[10^{23}]{Wcm^{-2}}$\cite{yoon2021}. Several studies have explored the possibility of creating matter from pure light in sufficiently strong laser fields\cite{bell2008,bulanov2010,ridgers2012}. In the centre-of-mass frame of this many-photon interaction, the field strength approaches the Schwinger field \cite{schwinger1951}, enabling the production of positron-electron pairs in vacuum through the non-linear Breit-Wheeler process \cite{breit1934}. Previous experiments have been able to reach this regime using a $\unit[47]{GeV}$ electron beam from SLAC colliding with an $\unit[0.4]{TW}$ laser pulse to produce around $100$ positron-electron pairs from around $10^{10}$ electrons \cite{burke1997}. This first produces bright $\gamma$-rays from inverse Compton scattering \cite{bula1996}, before the $\gamma$-rays themselves interact with the laser field. Another approach has been to use the strong atomic fields present within periodic crystal lattices \cite{esberg2010,nielsen2023}, with a $\unit[200]{GeV}$ electron beam driving trident pair production in a germanium crystal. Now, several laser facilities aim to utilise unprecedented laser powers and produce positron-electron pairs in vacuum by using laser-driven plasma wakefield accelerators\cite{nees2020, piazza2022}. 

Conducting these experiments in practice is significantly more complicated than may be imagined from idealised simulations, with the interactions requiring femtosecond- and micron-level synchronisation and alignment between the laser and electron beams. In order to successfully produce positron-electron pairs in vacuum in laser-driven experiments, this spatial and temporal jitter must be accounted for, typically by running many simulations over the range of possible configurations. However, running this many simulations is often computationally expensive, especially when the chances of producing a positron-electron pair can be vanishingly small with current-generation laser facilities. In order to determine the best experimental parameters from a large domain of options, it is necessary to increase the efficiency both of calculating the pair production rates, and of exploring a multi-dimensional parameter space.

Several Monte-Carlo and Particle-In-Cell codes have already increased computational efficiency by introducing particle merging \cite{timokhin2010,vranic2015,dong2024} and splitting \cite{kawrakow2004,ivanchenko2014,ramos-mendez2017,smets2021} algorithms. Merging techniques are useful when lots of particles are being produced, using too much computational memory, whereas splitting techniques are useful when few particles are produced, allowing codes to explore sparse areas of the parameter space and reduce the variance in the results. With Breit-Wheeler pair production a rare process with current laser parameters, particle splitting is a promising approach. However, most particle splitting algorithms such as Refs.~\cite{ivanchenko2014,smets2021} split the primary particle (in our case photons), which we generally have many of already, unnecessarily increasing computation in the particle push or tracking. On the other hand, algorithms like Refs.~\cite{kawrakow2004,ramos-mendez2017} split the secondary particle (in our case a positron-electron pair) only after it has been produced, which is unhelpful if the chances of it being produced in the first place are so small. We therefore want to perform particle splitting during the pair production step, which is a novel and computationally efficient approach but risks affecting the algorithm's accuracy.

Here, we find optimum parameters for electron-positron pair production in all-optical experiments by employing two techniques for performing many simulations with higher computational efficiency. First, we describe a particle splitting algorithm that performs the split during pair production, and show how it requires changes to the way we model pair production. This accurately calculates the rate of pair production for rare events without significantly increasing the runtime, allowing us to rapidly conduct many simulated experiments, including the effect of jitter. Second, we use a Bayesian optimisation scheme built on Gaussian process regression to efficiently explore the parameter space and find the maximum rate of pair production. This scheme is significantly faster than a brute force grid search and builds a fast surrogate model for estimating the rate of pair production across the domain. We show that more positron-electron pairs can be produced by combining the tightest possible laser focus with a few-cm-scale stand-off distance between the laser wakefield accelerator and the interaction point. This ameliorates the effect of jitter while retaining the high field strength required for pair production.

\section{Pair Production and Particle Splitting} \label{sec: Splitting}

\subsection{Pair Production Algorithm} \label{sec: Algorithms}

In both Monte-Carlo and Particle-In-Cell (PIC) codes, the emission of high energy photons and the production of positron-electron pairs are treated as discrete random and independent events which can occur in any fixed time interval. As such, they can be modelled using a Poisson distribution, where the average number of emissions per particle in a given timestep $\delta t$ are given by:
\begin{align}
\lambda_\gamma &= \frac{dN_\gamma(\chi_e)}{dt} \delta t \\
\lambda_{e^+,e^-} &= \frac{dN_{e^+,e^-}(\chi_\gamma)}{dt} \delta t,
\end{align}
where the electron and photon quantum parameters are given by ${\chi_e = |F_{\mu\nu} p_e^\nu|/m_eE_s}$ and ${\chi_\gamma = |F_{\mu\nu} \hbar k_\gamma^\nu|/m_eE_s}$, for a Schwinger field ${E_s= m_e^2c^3/e\hbar}$. The probability of emitting $n$ particles in a given timestep is then 
\begin{equation}
P(n|\lambda) = \frac{\lambda^ne^{-\lambda}}{n!}
\end{equation}

Often, in order to reduce the computational expense of sampling from a Poisson distribution at every timestep, codes use an optical depth approach (for instance in the EPOCH PIC code \cite{arber2015,ridgers2014}). If the emission rate is sufficiently low, the chances of two or more emissions in a single timestep is negligible and the probability of a single particle being emitted on a timestep before $t_N$ becomes:
\begin{align*}
P_\mathrm{emission}(t_0<t<t_{N}) &\approx 1 - P_\mathrm{no~emission}(t_0<t<t_{N})\\
&\approx  1 - \exp \left[ -\sum^N_{n=0} \lambda(t_n) \right].
\end{align*}
Now a uniformly distributed random number in the range $0\leq r<1$ is chosen to represent $P(t)$ and an emission occurs on the first timestep when $1 - \exp \left[ -\sum_n \lambda(t_n) \right] \leq r$, which is when the accumulated optical depth reaches $\sum_n\lambda(t_n) \geq -\ln(1 - r)$. Only one random number must be chosen for each emission event, calculated the moment a particle is created or the moment after an emission takes place, and a single running total tracks the optical depth. The optical depth method therefore efficiently models rare emissions which occur singly and consecutively.

This model is particularly relevant for pair production, as a photon ceases to exist after emitting a positron-electron pair, so more than one emission is impossible. In fact, emitting a photon also changes an electron's trajectory, due to radiation reaction, and so emitting multiple photons in a single timestep using the same value for $\chi_e$ is also unphysical if the photon energy is sufficiently high. The optical depth method is therefore generally very well suited for the problems we are modelling.

However, if emissions are very rare, such as for Breit-Wheeler pair production, a tremendously large number of particles must be used to model the likelihood of even a single emission, with the vast majority of these particles not participating in the interaction at all. Each of the particles must be tracked through the simulation even if they do not contribute at all to the creation of pairs. In this situation, despite the strengths of the optical depth method, modelling single rare emissions is extremely computationally inefficient.

\subsection{Particle Splitting Algorithms}

We therefore want to introduce particle splitting into the calculation of Breit-Wheeler pair production in Monte-Carlo and PIC codes, during the pair production step rather than before or afterwards. In PIC codes, each macroparticle represents a large number of underlying real particles, which move around the simulation grid together and sample the underlying distribution in $(\mathbf{x},\mathbf{p})$ space with a given weight $w$. When PIC codes model random emission processes they typically use the method discussed in Section~\ref{sec: Algorithms}, sampling the probability distribution function once per timestep and creating child particles which also each have a weight of $w$. Here we describe how we can let each macroparticle move around as before, but instead independently sample the chances of producing a positron-electron pair many times per timestep at a splitting number $uw$, making the average number of emissions per timestep now equal to:
\begin{equation}
\lambda_{e^+,e^-} \rightarrow uw \frac{dN_{e^+,e^-}(\chi_\gamma)}{dt} \delta t.
\end{equation}
The weight of any pairs produced will be reduced proportionately, from $w$ to $1/u$, and the weight of the photon will be slightly reduced to account for the decay, from $w$ to $w-1/u$. If this weight falls to zero the photon is removed from the simulation. Particle splitting does not affect the number of parent particles and does not significantly increase the cost of the particle push, only affecting the number of pair particles. In this way, emission events can be made more common without greatly increasing the runtime, more efficiently modelling pair production.

However, there is a risk to naively using the optical depth algorithm with a high rate of particle splitting for this purpose. Firstly, if the degree of particle splitting $u$ is sufficiently high that $\lambda_{e^+,e^-} \sim 1$, there is now a non-negligible chance that more than one pair will be produced per timestep, giving $P_\mathrm{emission}(0<t<t_{N}) \neq  1 - \exp \left[ -\sum^N_{n=1} \lambda(t_n) \right]$. Secondly, even when only one or fewer pairs is produced in a timestep, the optical depth is refreshed after each emission using a new random number. This wipes the history for all particles represented by the macroparticle, not just the fraction $1/u$ that has been emitted, making emissions no longer independent. Without particle splitting, neither of these are issues, as photons are killed immediately upon producing just one electron-positron pair. Using particle splitting while using a single optical depth per particle, however, can give a drastically different estimate for the rate of pair production.

This problem can be resolved in three different ways, with different degrees of accuracy:

\emph{I} - Sampling the full Poisson distribution on every timestep to model particle splitting. The average number of emissions in each timestep is still increased by a factor $u$ but now any number of emissions are possible in any timestep. If at least one emission occurs in a single timestep, $n>0$, a single pair of electron and positron macroparticles can be produced, each with a weight $n/u$. Alternatively, if a larger number of pair particles are desired, for instance to better sample an energy distribution, $n$ macroparticles can be produced, each with a weight $1/u$. In either case, the weight of the photon is reduced to $w-n/u$. This is a relatively simple extension for codes like PTARMIGAN \cite{blackburn2023}, which already calculate the emission probability on every timestep. It ensures that pair production events are independent and conserves energy on every emission, while allowing for an arbitrarily large degree of particle splitting.

\emph{II} - Adding the new random optical depth onto the previous value, including the `overshoot' by which the optical depth fell below zero. Another emission occurs when this additive optical depth falls below 0 on a subsequent timestep. As the optical depth becomes the sum of all the random numbers $-\sum_i\ln(1-r_i)$, minus the total accumulated event path length $\sum_n\lambda t_n$, this maintains both the history of the particle and the independence of the emissions, while using only a single random number for each emission event. In this case, the parent particle weight is not updated after each emission, in order to make sure the emissions are still independent. Instead, the particle is killed with a probability of $1/u$, such that at higher splitting rates the majority of particles are unaffected by emissions. This is a natural extension for codes that use optical depths, like EPOCH \cite{ridgers2014} or Smilei \cite{derouillat2018}. It makes pair production events independent again while conserving energy on average.

\emph{III} - `Sub-cycling' the addition of multiple random optical depths onto the previous value to represent multiple emissions. Typically the `additive optical depth' approach will not allow for more than one emission per timestep, so will also fail for high rates of pair production $\lambda \sim 1$, where more than one pair can be produced by a single macroparticle per timestep. However, it's possible to repeat the process multiple times, continuing to compute emissions so long as the optical depth remains below zero. In practice, this repeated optical depth approach is the technique used by many Poisson sampling algorithms when the mean is not much greater than one, and so if $\lambda$ is around one or less this sub-cyling approach is still more efficient than Poisson sampling. However, sub-cycling still does not update the parent particle weight after each emission and so only conserves energy on the ensemble average.

The four approaches are summarised in Table~\ref{tab: algorithms}. The most expensive and most accurate approach is full Poisson sampling, which allows for multiple emissions per timestep at very high field strengths and conserves energy on every emission, but requires calculating a random number every timestep (more than one random number if there is more than one emission) and is therefore more expensive. It also requires modelling particles with a wide range of different weights ($w - n/u$ and $n/u$ for any integer $n$), which can add computational challenges. The naive optical depth approach and additive optical depth approaches are equally computationally cheap, but using an additive optical depth extends the range where upscaling is valid to areas where a particle will have more than one emission over the course of the simulation, at the cost of only conserving energy on average. Subcycling extends the range of applicability still further by allowing multiple emissions per timestep, at the cost of more computation only when multiple emissions per timestep are required. From here on we will only use the version of the additive optical depth with sub-cycling included. All of these optical depth approaches have simpler particle weights, either `big' macroparticles with weight $w$ or `small' macroparticles with weight $1/u$. This makes subsequent numerical behaviour simpler, such as by ignoring `small' particles in subsequent collisions.

\begin{table}
\hspace{-2cm}
\begin{tabular}{|c|c|c|c|c|c|}
\hline
\emph{Algorithm} & \emph{Computation} & \emph{Valid when:} & \emph{Simple Weights?} & \emph{Conserves energy?} \\
\hline
Naive optical depth & Min($\lambda$,$1$) & $\lambda\ll 1$ & Yes & Exactly \\
Additive optical depth & Min($\lambda$,$1$) & $\lambda \lessapprox 1$ & Yes & On average \\
Additive plus subcycling & $\lambda$ & Any $\lambda$ & Yes & On average \\
Poisson & Max($\lambda$,$1$) & Any $\lambda$ & No & Exactly \\
\hline
\end{tabular}
\caption{Summary of the different particle splitting techniques for an average number of pairs produced per timestep $\lambda=uw\frac{dN_+}{dt}\delta t$, including a particle splitting rate $u$. Columns describe the approximate computational cost (as the average number of random numbers required per macroparticle per timestep), the range of $\lambda$ where the algorithm is appropriate, whether the algorithm keeps weights simple (either $w$ or $1/u$), and the energy conservation of the algorithm. }
\label{tab: algorithms}
\end{table}

\subsection{Particle Splitting Simulations}

\begin{figure}
\includegraphics[trim={3cm 12cm 0 13cm},scale=1.0]{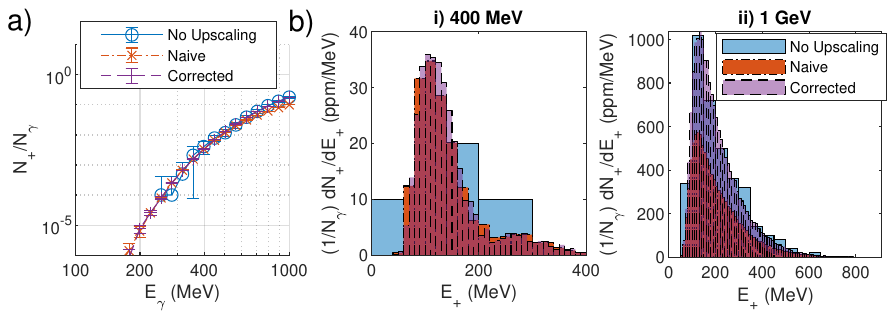}
\caption{a) Rate of electron-positron pair production in a tightly focussed 1 PW beam against photon energy, comparing results without splitting (blue circles), with naive splitting (red 'x's), with Poisson-sampling splitting (purple crosses), and with additive optical depths (green diamonds). b) Positron spectra predicted without splitting (blue) and with naive (red) and corrected splitting (purple), for photons at 400 MeV (i) and 1 GeV (ii).}
\label{fig: upscaling motivation}
\end{figure}

We tested the three techniques for particle splitting in Monte-Carlo simulations of a head-on collision between high energy photons and an intense laser pulse using the \texttt{QED-Cascade} code \cite{watt2021,watt2025}. A $\unit[1]{PW}$ pulse with $\unit[25]{J}$ in $\unit[25]{fs}$ was simulated, focussed to a spot waist of $w_0=\unit[2]{\mu m}$ and a peak intensity of $\unit[2.2\times10^{22}]{Wcm^{-2}}$ or $a_0\approx100$. $1000$ photon particles, each with an initial weight of $w=1$, were collided with the laser pulse at an angle of $15^\circ$, intersecting the peak laser intensity at best focus. The photon beam was a mono-energetic and divergence-free point source with energies varying from $\unit[100]{MeV}$ up to $\unit[1]{GeV}$. 10 simulations were run at each photon energy, with the results shown in Fig.~\ref{fig: upscaling motivation}. This shows both the number of pairs produced and the change in the predicted energy spectrum of the positrons.

\begin{figure}
\includegraphics[trim={3cm 12cm 0 13cm},scale=1.0]{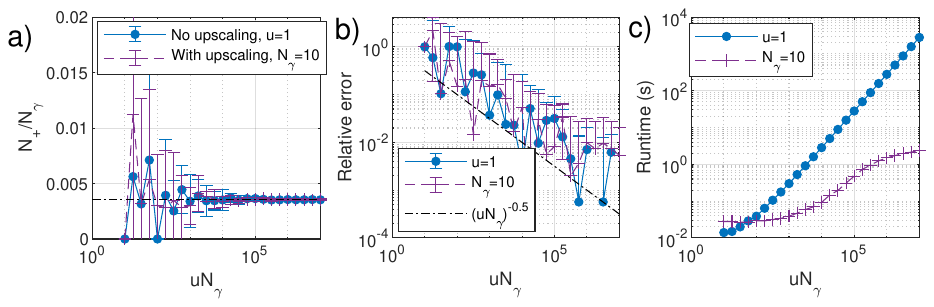}
\caption{a) Rate of pair production for 400 MeV photons in a 1 PW beam against the number of photon particles $N_\gamma$ and the rate of splitting $u$, showing results with no splitting ($u=1$) and with a fixed number of particles but variable splitting ($N_\gamma=10$). b) Accuracy of the estimated rate of pair production plotted against $uN_\gamma$. c) Runtime of the simulation plotted against $uN_\gamma$}
\label{fig: upscaling accuracy}
\end{figure}

At photon energies below $\unit[300]{MeV}$, the rate of pair production is sufficiently small (below $10^{-4}$ positrons per photon) that the original simulations without splitting do not produce any positron-electron pairs, leaving the simulations unable to capture the Breit-Wheeler process. Even above this, very small numbers of pairs are produced, giving significant variation from shot noise, as shown by large error bars at each point. At $\unit[400]{MeV}$, only four positron macroparticles were produced, giving large shot noise and making it hard to resolve the positron energy distribution. For convergence testing at these low energies we require simulations with more photon macroparticles, described later (see Fig.~\ref{fig: upscaling accuracy}).

A naive attempt to increase the number of positrons produced simply increased the rate of pair production and reduced the weight of the pair macroparticles produced. This can capture the Breit-Wheeler process at much lower pair production rates, down to $10^{-7}$ pairs per photon in this case. For photon energies between $300$ and $\unit[600]{MeV}$ this naive splitting gives results that agree well with the original simulation but with many more positron macroparticles and hence much lower shot noise (for instance producing 3400 macroparticles at $\unit[400]{MeV}$). However, above $\unit[600]{MeV}$, the naive splitting underestimates the rate of pair production by a factor of around two. This is shown most clearly in the predicted positron energy distribution at $\unit[1]{GeV}$, demonstrating that the naive splitting is not reliable.

Using both of the corrected splitting methods, returning to sampling the Poisson distribution and adding new optical depths onto the cumulative total, the simulations are able to accurately resolve the pair production rate both at photon energies as low as $\unit[150]{MeV}$, and up to $\unit[1]{GeV}$. Here, the shot noise is lower and the larger number of positron macroparticles (100,000 at $\unit[1]{GeV}$) enables fine resolution of the pair production rates and the positron energy distributions. Both corrected splitting methods are therefore effective for simulating Breit-Wheeler pair production across a wider range of photon energies, even with a relatively small number of photon macroparticles. For the rest of the paper, we focus on the Poisson approach, which is closer to the true physical picture at the cost of a higher runtime.

We can compare the accuracy of simulations with and without splitting by running simulations at a range of splitting rates $u$ and number of photons $N_\gamma$ for a fixed photon energy of $\unit[400]{MeV}$, again colliding with a laser pulse of $\unit[25]{J}$ in $\unit[25]{fs}$. Increasing either $u$ or $N_\gamma$ should increase the number of positron macroparticles produced and hence improve the estimate of the underlying pair production rate, but will also increase the runtime.

Initially, we run simulations with no splitting ($u=1$), but with the number of photon macroparticles varying from just $N_\gamma=10$ up to 10 million; the results are shown in Fig.~\ref{fig: upscaling accuracy}. As the number of photon macroparticles increases, the rate of pair production predicted by the simulations converges to a consistent value, at $(3.53\pm0.02)\times 10^{-3}$ positrons per photon. Both the error on the rate of pair production and the variation across simulations drop with the increasing number of photon macroparticles, limited by the shot noise as expected ($1/\sqrt{N_+}\propto1/\sqrt{uN_\gamma}$) until the relative error is below $1\%$. The runtime, however, increases to over 2,800 seconds per simulation, meaning the set of 10 simulations takes almost 8 hours. The increase in accuracy therefore comes at a significant cost.

Keeping the number of macroparticles fixed at $N_\gamma=10$ and increasing splitting, on the other hand, increases the accuracy of the prediction while having much less effect on the runtime. The rate of pair production converges to the same value of $(3.55\pm0.05)\times 10^{-3}$ positrons per photon and the accuracy improves at the same rate, $\propto 1/\sqrt{uN_\gamma}$. The runtime, however, only reaches 2.3 seconds per simulation, three orders of magnitude faster for the same accuracy, meaning that using particle splitting is an efficient way of accurately simulating Breit-Wheeler pair production.

\section{Monte-Carlo Simulations of Realistic Experiments}

\subsection{Simulating Shot-to-Shot Jitter} \label{sec: Eg Sims}

Running many simulations quickly is particularly useful when optimising parameters for real laser experiments, where there is significant random `jitter' in experimental parameters from shot to shot that reduce the peak quantum parameter of the collision. Even if the energy of the electron beam and the colliding laser are very stable, changes in the laser pointing often cause the electron beam position to move by tens of microns between shots. Similarly, even if the synchronisation of the two laser beams is controlled to a few parts per million, tens of femtosecond shifts in timing are common. When the focus of the colliding beam is only microns across and a few tens of femtoseconds in duration, these effects reduce the measured rate of pair production by orders of magnitude relative to the best-case head-on collision. Simulations using the ideal parameters are often unrealistically optimistic about what measurements are possible.

Now we have demonstrated that particle splitting gives significant reductions in runtime for the same level of accuracy, we are able to run many consecutive Monte-Carlo simulations to model the effect of jitter and capture pair production even at low laser intensity. By modelling hundreds or thousands of consecutive laser shots with the experimental parameters randomly drawn from the known distributions, we can better simulate a realistic laser experiment and estimate the total accumulated rate of pair production. Note, for benchmarking purposes we conduct most of these simulations with mono-energetic electrons, with zero energy spread and no jitter in the electron energy.

\begin{figure}
\includegraphics[scale=0.45,trim=0cm 4cm 9cm 2cm]{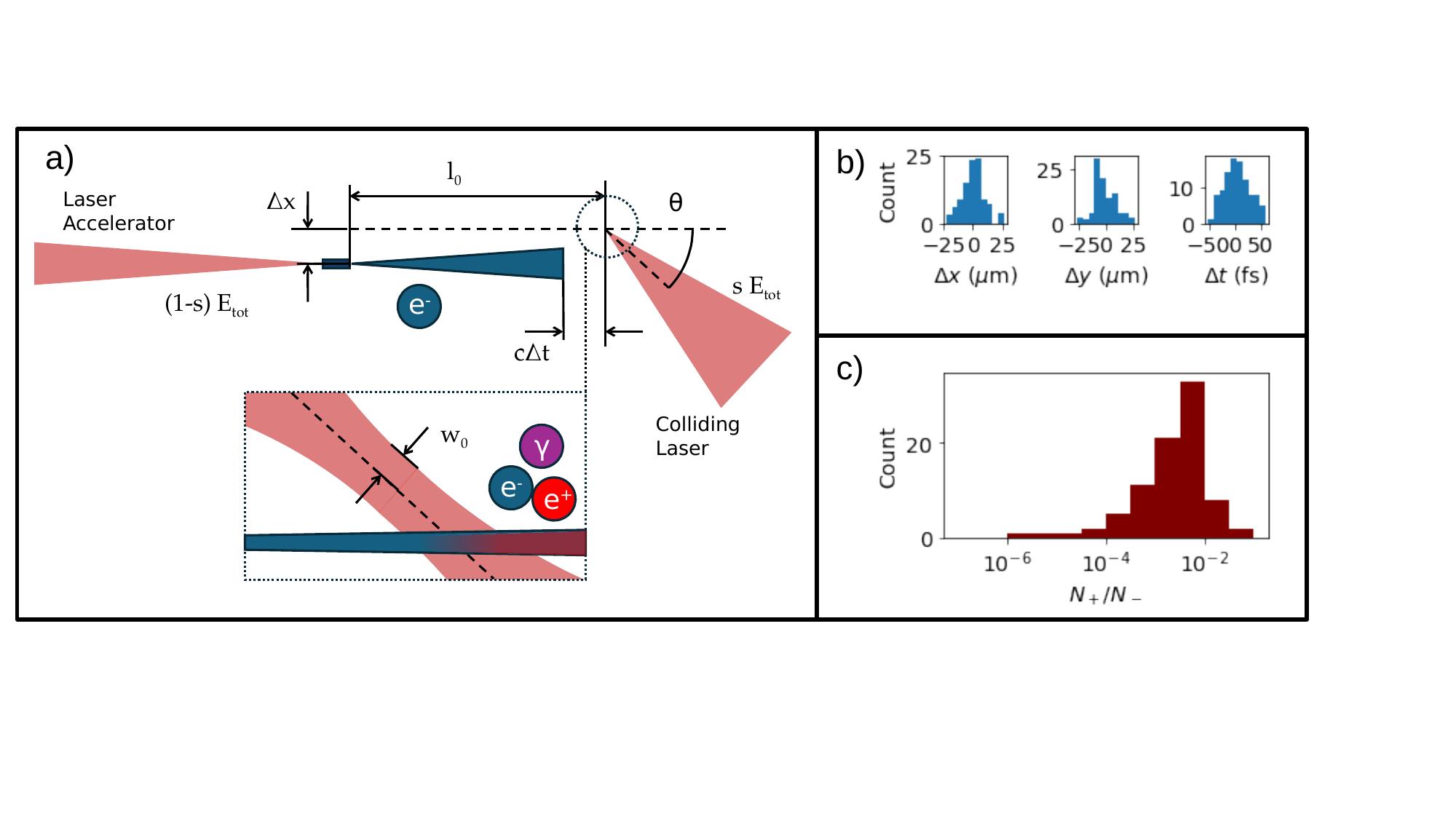}
\caption{a) Set up of the simulated experiment, with a laser pulse driving an electron beam from a laser wakefield accelerator, which then collides with an intense laser pulse at an angle of $\theta$. A stand-off distance of $l_0$ allows the electron beam to expand before it collides with the laser pulse. The colliding laser pulse is focussed to a spot size of $w_0$ and contains $s E_\mathrm{tot}$ of energy, compared to $(1-s)E_\mathrm{tot}$ in the accelerator beam. There is a random offset in both space and time between the peak intensity at the focus of the laser pulse and the electron beam. b) Prior distributions of normally distributed offsets $\Delta x$, $\Delta y$, and $\Delta t$ over 100 simulated shots. c) The corresponding posterior distribution of positron production rate for $\theta=15^\circ$, $l_0=\unit[10]{cm}$, $w_0=\unit[2]{\mu m}$, and $\unit[7.5]{GeV}$ electrons colliding with $sE_\mathrm{tot}=\unit[25]{J}$.}
\label{fig: simulated experiment}
\end{figure}

An example of such a simulated experiment is shown in Fig.~\ref{fig: simulated experiment}. An initial electron beam with an energy of $\unit[7.5]{GeV}$ collides with an intense laser pulse, again with $\unit[25]{J}$ in $\unit[25]{fs}$ focussed to a $\unit[2]{\mu m}$ spot. The colliding laser parameters are chosen to reflect the designs of the ZEUS and CoReLS laser facilities \cite{yoon2021,willingale2023} (see also the APOLLON laser \cite{zou2015}), while these high electron energies have recently been achieved in guided laser wakefield acceleration experiments \cite{gonsalves2019,miao2022,picksley2024}. Now the prior distributions of laser pointing $\Delta x$ and $\Delta y$ (both $\sim \mathcal{N}(0, \unit[10]{\mu m})$) and timing jitter $\Delta t$ $\sim \mathcal{N}(0, \unit[25]{fs})$ lead to imperfect collisions with a distribution of $\chi_e$ and $\chi_\gamma$. The posterior distribution for the number of pairs produced $N_+/N_-$ includes both an error from shot noise and a much more substantial variation from changes in jitter. This simulated experiment took 500 seconds to run on a single core, giving an average rate of pair production of $N_+/N_-\approx0.013\pm0.001$. This is two orders of magnitude lower than for the ideal `head-on' collision with $\Delta x=\Delta y=\Delta t=0$, which gives a positron rate of $N_+/N_-\approx1.2$.

\subsection{Optimising Experimental Conditions} \label{sec: Eg optimising}

We can now use simulated experiments to optimise the rate of pair production in a laser collider set-up using Gaussian Process Regression. By running many Monte-Carlo simulations at each set of nominal experimental parameters -- with variations around those conditions randomly drawn from given distributions -- we can determine what parameters to aim for when trying to measure Breit-Wheeler pair production in a real laser experiment. This is a noisy function over a multi-dimensional parameter space, with significant errors from both jitter in the experimental parameters and stochastic noise in the rate of pair production. A Bayesian optimisation approach using Gaussian Process Regression is therefore well suited to rapidly exploring the parameter space and then focussing measurements on the region with the best chances of producing high numbers of pairs.

A Bayesian optimisation algorithm attempts to efficiently maximise a given output from a noisy function, in this case the number of pairs produced in a Monte-Carlo simulation. It does so by building a surrogate model with an associated error, which acts as a proxy for the underlying real function while being much faster to evaluate. In this case, for the surrogate model we use a Gaussian process, which is a stochastic process with a normally distributed error. This continuously interpolates between measurements in the multi-dimensional space so as to ensure that the surrogate model is consistent with the data points and the specified error.

Every time the algorithm makes a new measurement it uses the surrogate model to calculate an acquisition function, takes a measurement where the acquisition function is highest, and uses the results of these measurements to update the surrogate model with the new information. The acquisition function we used is the Expected Improvement, which integrates over the error distribution of the surrogate model to find the place in parameter space where the expected increase from the current maximum to the surrogate model output is highest. This includes locations both where the surrogate model is high, and also where the error on the surrogate model is high, reflecting our lack of confidence in the surrogate to accurately represent the underlying function. Over many measurements, the surrogate model becomes an increasingly accurate proxy for the underlying function, the error on the surrogate model drops, and the acquisition function predicts smaller improvements.

To begin with, we optimised a simulated experiment with just a single variable, the stand-off distance $l_0$, for a fixed electron beam divergence of $\unit[1]{mrad}$ FWHM. The larger the stand-off distance, the larger the electron beam at the collision point, presenting a bigger target to hit. However, a larger electron beam also means a lower electron density and fewer electrons participating in the collision when one occurs. These simulations used $\unit[7.5]{GeV}$ electrons colliding with $E_\mathrm{coll}=\unit[25]{J}$ of laser energy at an angle of $\theta=15^\circ$, where the colliding laser pulse was focussed to $w_0=\unit[2]{\mu m}$ and had a duration of $\unit[25]{fs}$ FWHM, giving a peak power of around 1 PW.

\begin{figure}
\includegraphics[scale=1.0,trim=4cm 12cm 0cm 13cm]{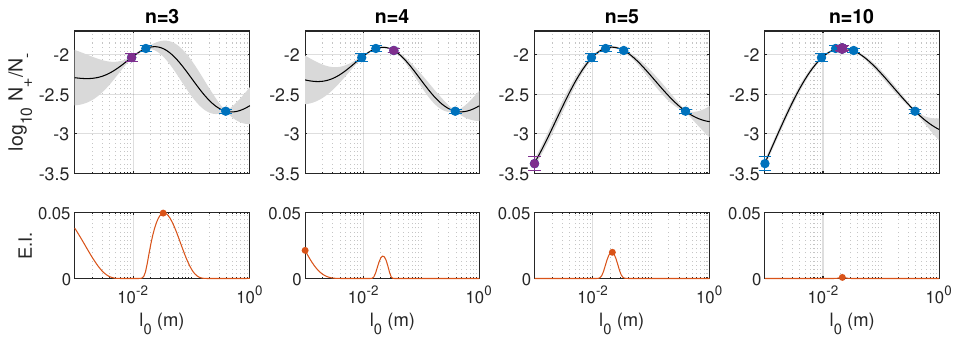}
\caption{Top panels) Positron production at different stand-off distances $l_0$ predicted by simulated experiments, overlaid on the surrogate model produced by Gaussian Process Regression. Circles show the mean of $log_{10}(N_+/N_-)$ while the error bars show the standard error over 100 simulated shots, including $\unit[10]{\mu m}$ spatial jitter and $\unit[25]{fs}$ temporal jitter. The model's predictions after each number of points $n$ are shown by the black line and the standard deviation on the model is shown by the shaded region. New points are shown in purple while previous points are shown in blue. Bottom panels) The expected improvement acquisition function predicted by the model against the stand-off distance. Peaks in the expected improvement are marked with dots and are used to calculate the next location to sample at $n+1$.}
\label{fig: l0 example}
\end{figure}

The results from the GPR algorithm are shown in the top panels of Fig.~\ref{fig: l0 example} as more simulated experiments are run at different values of the stand-off distance $l_0$. Initially, after just $n=3$ initial random points, the model carries a high uncertainty across most of the domain. As more data points are taken this uncertainty gets smaller until by $n=5$ the underlying function near the maximum and the position of the maximum are known accurately, while uncertainty remains elsewhere in the domain. Further points up to $n=10$ continue to reduce the error near the maximum. After each simulation, the next set of simulated experiments are run where the expected improvement (shown by the bottom panels) is high, narrowing down the possibilities. As more points are sampled and the confidence of the model increases, the expected improvement falls. With the `explore' parameter set to just $0.2$, the algorithm increases the expected improvement in the region close to the current maximum rather than in other regions of the domain. Ultimately, this gives a maximum of $0.015\pm0.001$ positrons per initial electron at a stand-off distance of $l_0=\unit[17]{mm}$.

\begin{figure}
\includegraphics[scale=1.0,trim=4cm 11cm 0 13cm]{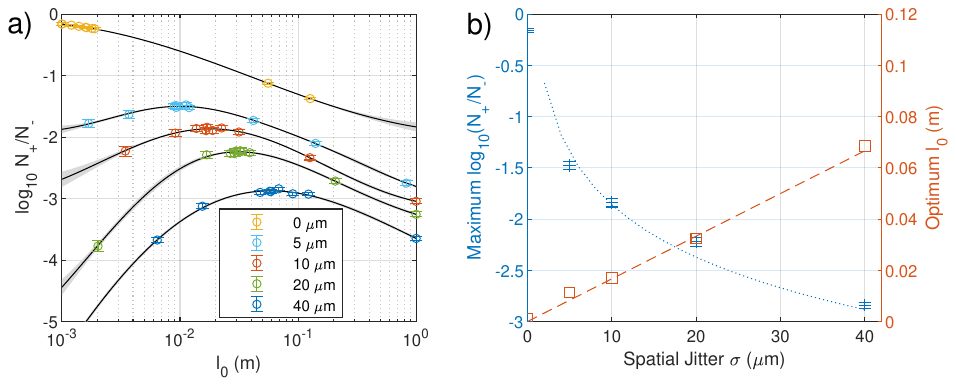}
\caption{a) Gaussian process optimisation of the rate of pair production at different levels of spatial jitter, from $\unit[0]{\mu m}$ standard deviation up to $\unit[40]{\mu m}$ standard deviation. 100 simulations were run at each point in parameter space, with the mean and error shown by the coloured points. In each case, the results were fitted to a Gaussian Process surrogate model, with the mean shown by the black lines and an error shown by the shaded region. b) Summary of the maximum rate of pair production (left axis, shown in blue) and the optimal stand-off distance (right axis, shown in red), plotted against the level of spatial jitter. The blue dotted trend line shows a power law fit $N_+/N_- = 0.7 \sigma^{-1.7}$ and the orange dashed trend line shows a linear fit $l_0 = \sigma / \unit[0.6]{mrad}$ respectively, for a spatial jitter given by $\sigma$.}
\label{fig: jitter results}
\end{figure}

We can also look at the impact of changing the other experimental conditions, such as the jitter, with examples shown in Fig.~\ref{fig: jitter results}. With no spatial jitter the maximum rate of pair production is $0.69\pm0.03$ pairs produced per initial electron, with zero stand-off-distance, reflecting the high electron energy and laser intensity. However, increasing the spatial jitter causes a reduction in the average positron production rate by orders of magnitude, as the likelihood of an electron hitting the most intense part of the laser at a given stand-off distance reduces as $\sigma^{-2}$ for a spatial jitter given by $\sigma$. This effect can be ameliorated by increasing the stand-off distance and hence the size of the electron beam at the collision point, with the optimal stand-off distance increasing to $\unit[69]{mm}$ at $\sigma=\unit[40]{\mu m}$. At these large stand-off distances, the larger electron beam is easier to hit and the increase in the number of shots that hit the electron beam outweighs the disadvantage of only a small fraction of the electron beam interacting with the laser pulse on each shot. At higher spatial jitters the optimum stand-off distance increases linearly such that the optimal size of the electron beam should be comparable to the spatial jitter.

\subsection{Optimum parameters for Non-Linear Breit-Wheeler pair production}

Now we have demonstrated a one-dimensional optimisation, we will use Gaussian Process Regression to maximise the number of positrons produced across several parameters in simulated experiments. To find the optimum parameters, we consider changes in the energy split $s=E_\mathrm{coll}/E_\mathrm{tot}$, for $E_\mathrm{tot}=\unit[100]{J}$. The resulting electron energy is assumed to be linear with laser energy as $E_e = \unit[10(1-s)]{GeV}$. This scaling reflects all-optical inverse Compton scattering experiments over the last 13 years at a range of facilities \cite{tsai2016,taphuoc2012,chen2013,poder2018,mirzaie2024}. We also vary the laser waist $w_0$ between $\unit[2]{\mu m}$ and $\unit[20]{\mu m}$, and the stand-off distance $l_0$ between $\unit[1]{mm}$ and $\unit[1]{m}$. A smaller laser spot and shorter duration means the laser reaches a higher intensity, enabling larger quantum parameters $\chi_e$ and $\chi_\gamma$. However, it also provides a smaller target such that when the spatial or temporal jitter is large, the electron beam is more likely to miss the laser pulse completely and encounter only a very low quantum parameter. Similarly, a smaller stand-off distance means the electron bunch is small at the collision point, with almost all electrons encountering a high laser intensity, but it increases the probability of the electron beam missing the colliding laser entirely. Larger stand-off distances increase the probability of a collision but reduce the number of electrons taking part in the interaction. We use a particle splitting rate of $u=1000$ to better resolve the positron production rate in lower field regions of the parameter space and reduce the error from shot noise.

Until now, we have performed the simulations for monoenergetic photons (section~\ref{sec: Splitting}) and monoenergetic electrons (sections~\ref{sec: Eg Sims} and \ref{sec: Eg optimising}). In practice, however, LWFA electron beams typically have high variability shot-to-shot and significant energy spread. For these simulations we model energy spread by sampling electrons from a Gaussian distribution with a standard deviation $10\%$ of the electron mean energy. Shot-to-shot variation in the electron mean energy itself is modelled by sampling from a Gaussian distribution with a mean of $\unit[10(1-s)]{GeV}$ and a standard deviation of $\unit[(1-s)]{GeV}$, where $s=E_\mathrm{coll}/E_{tot}$ as before. This reflects approximately $10\%$ shot-to-shot variation in electron energy and $10\%$ energy spread, as in e.g. Refs.~\cite{picksley2024,v.grafenstein2023}. The electron beam divergence remains fixed at $\unit[1]{mrad}$. These parameters are representative of all-optical collision experiments, but are not tied to a particular setup or facility.

\begin{figure}
\includegraphics[scale=1,trim=1cm 0 0 0]{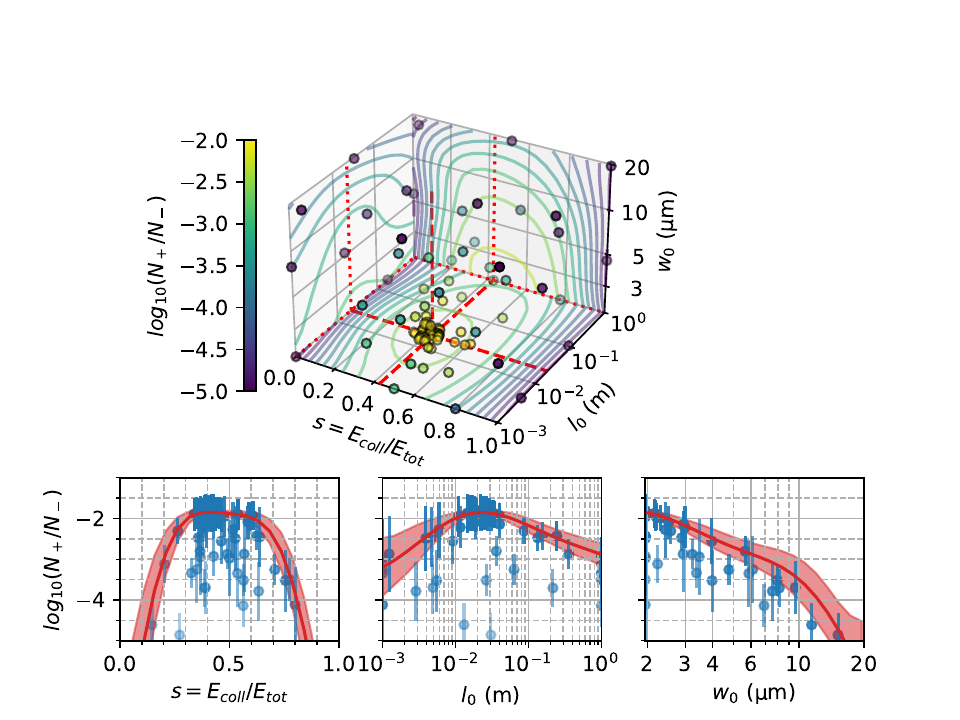}
\caption{Bayesian optimisation of the number of positron-electron pairs produced as an electron beam passes through an intense laser pulse in three dimensions. Top) A 3D scatter plot of the simulated experiments, coloured by positron yield. The results are fitted to a surrogate model, which is shown as contours, each showing the model predictions averaged over one axis. Below) The results of simulated experiments are plotted against each axis in turn and compared to the surrogate model mean (red line) and $2\sigma$ error (shaded area). The model predictions are calculated along lines through the optimum experimental parameters (shown as red dashed lines in the scatter plot). The transparency of the data points reflects the distance from this line, with lighter points further away from the optimum in parameter space. Error bars on the data points show the standard deviation over the 100 simulated experiments.}
\label{fig: GPR Positrons}
\end{figure}

Figure~\ref{fig: GPR Positrons} shows the results of the simulations, and the Bayesian optimisation as a series of projections through the data. The maximum number of positrons obtained in these simulated experiments with $\unit[10]{\mu m}$ spatial jitter is ${(1.57\pm0.18)\times 10^{-2}}$, or around 2 pairs per 100 electrons. This optimum is obtained at an energy split of $s=0.43$, giving a mean electron energy of $\unit[5.7]{GeV}$ colliding with $\unit[1.7]{PW}$ of laser power. As in the previous section, the positron yield is maximised at a stand-off distance where the electron beam size is comparable to the $\unit[10]{\mu m}$ spatial jitter, here at $l_0=\unit[22]{mm}$. The minimum laser waist of $\unit[2]{\mu m}$ gives a peak field strength of $a_0 = 132$ and yields the largest number of positrons, with the rate of pair production falling four orders of magnitude with a factor of $10$ increase in $w_0$. This highlights the extremely strong dependence of Breit-Wheeler pair production on the quantum parameter, which here would nominally reach up to $\chi_e \approx 9$ for an ideal collision with no radiation reaction. These simulated experiments demonstrate that pair production requires the highest possible laser intensity at focus, even if it means accepting a relatively large stand-off distance, with few electrons reaching the peak field strength.

\subsection{Optimum parameters for measuring inverse Compton scatteriing}

We can also use simulated experiments to maximise the synchrotron radiation emitted by an electron beam as it collides with an intense laser pulse. This radiation is emitted at energies up to the 10s-100s of MeV scale and can contain a significant fraction of the electron energy. The high-energy gamma rays produced are a key result of the many-photon interactions in the strong field regime and can lead to the production of positron-electron pairs. Generally the energy in the radiation is increased by higher electron energies and strong field strengths (through the quantum parameter $\chi_e$), and a longer interaction time. With the significant jitter of these simulated experiments it is also important to maximise the number of particles involved in the interaction. We vary the same three parameters the energy split $s=E_\mathrm{coll}/E_\mathrm{tot}$, the stand-off distance $l_0$, and the laser waist $w_0$.

\begin{figure}
\includegraphics[scale=1,trim=1cm 0 0 0]{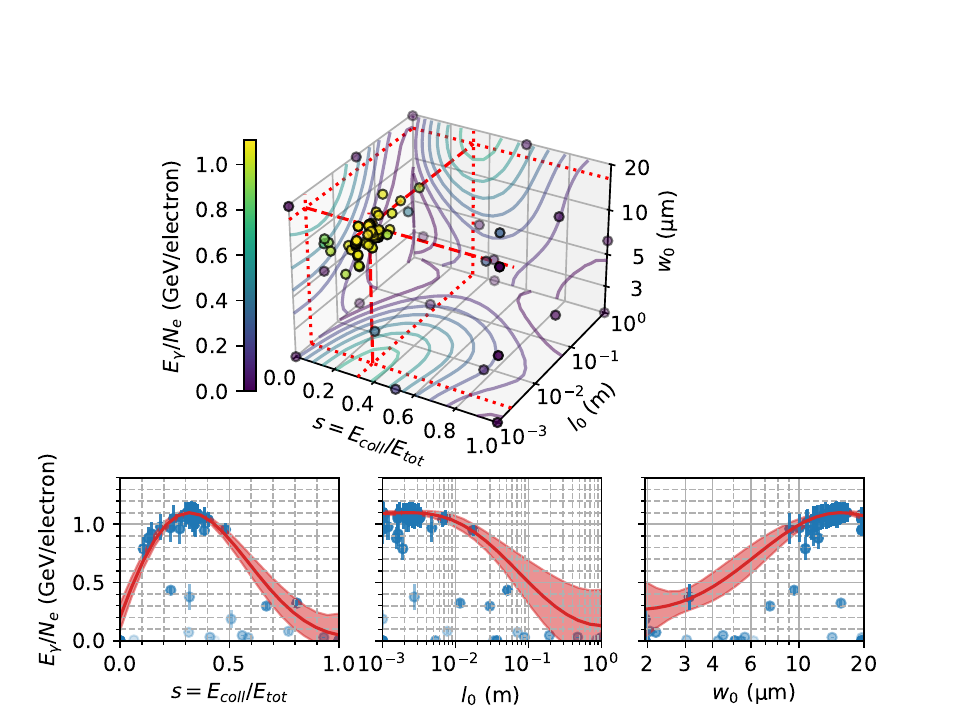}
\caption{The same plotting procedure as Fig.~\ref{fig: GPR Positrons} is now used to show the Bayesian optimisation of the radiation yield. Again, data points show the results of 100 simulated experiments taken at given points in parameter space, with the colour describing the radiation yield. The contours (top) show projections of the surrogate model averaged over each axis in turn, while the solid red lines and shaded regions (bottom) show the surrogate model mean and $2\sigma$ error taken along lineouts through the optimum point in parameter space. The error bars on the data points here describe the standard error on the mean radiation yield.}
\label{fig: GPR RR}
\end{figure}

Figure~\ref{fig: GPR RR} shows the result of the three-dimensional maximisation of the radiated energy. Points selected for simulated experiments are rapidly focussed in on the most promising region, giving a more accurate surrogate model with lower error bars near the maximum. This allows us to quickly determine the best parameters at which to run experiments to produce the maximum amount of radiation from an electron beam colliding with a laser pulse. In this situation, with \unit[100]{J} of laser energy available, the optimum is found with \unit[68]{J} used to accelerate the electron beam to \unit[6.8]{GeV}, and \unit[32]{J} used to collide with it. This matches the naive expectation of maximising the peak quantum parameter $\chi_e\propto\gamma\sqrt{I_0}\propto (1-s)\sqrt{s}$.

However, the laser waist which maximises the radiated energy is now around \unit[16]{$\mu$m}, much larger than the tightest possible spot and resulting in a peak intensity of $a_0\approx14$, much lower than the maximum achievable $a_0=200$. This increases the number of successful collisions by greatly increasing the volume around focus. When the laser waist is this large, the electron beam can be small and still collide effectively, such that the maximum energy radiated is again located at the minimum stand-off distance of $l_0=\unit[1]{mm}$.

\section{Conclusion}

In this work, we have demonstrated steps to efficiently optimise experimental conditions for all-optical Breit-Wheeler pair production using Monte-Carlo simulated experiments. First, we introduce two new algorithms for particle splitting which enables us to quickly and accurately sample the probability of producing a positron-electron pair many times within a single macroparticle. In contrast to the naive approach using a single optical depth, these techniques correctly predict the rate of pair production over at least an order of magnitude of variation in the quantum parameter $\chi_\gamma$, while giving runtimes up to a thousand times faster than by increasing the number of macroparticles. Secondly, this makes it practical to run many Monte-Carlo simulations at a single set of parameters in order to account for experimental jitter in laser pointing and timing. Given prior distributions for the shot-to-shot jitter, these simulated experiments predict a posterior distribution of rates of pair production, which gives the practical likelihood of measuring positron-electron pairs in experiments.

This approach makes it possible to efficiently optimise experimental parameters, maximising the likelihood of measuring Breit-Wheeler pair production in laser-driven experiments. Gaussian Process regression was used to build surrogate models of the rate of pair production and synchrotron radiation produced in simulated collisions between an LWFA electron beam and an intense laser pulse. These showed that laser pointing jitter can be ameliorated by increasing the size of the laser focal spot or by allowing the electron beam to expand after leaving the wakefield accelerator, with a stand-off distance between the accelerator and the collision point. Synchrotron radiation, which requires a lower laser intensity than pair production, was maximised with a larger focal spot ($\sim \unit[15]{\mu m}$), and no stand-off distance. In contrast, a stand-off distance of a few centimeters maximised the rate of pair production by increasing the size of the electron beam to tens of microns, comparable to the scale of the pointing jitter. This increases the number of electrons which collide with the laser focus, a focus which should be as small as possible (a few microns) in order to reach the highest possible laser intensity and quantum parameter. Together, these parameters enable near-term laser facilities to produce pairs at a rate of around one pair per one hundred electrons, even with realistic timing and pointing jitter.

\small{
This work was funded by EPSRC grant EP/V049461/1 and in part by the Air Force Office of Scientific Research (FA9550-24-1-0053). The Viking cluster was used during this project, which is a high performance compute facility provided by the University of York. We are grateful for computational support from the University of York IT Services and the Research IT team. Data is available upon reasonable request and code is available at \cite{arran2025}.
}

\bibliographystyle{iopart-num}
\providecommand{\newblock}{}

\end{document}